\documentclass[conference]{IEEEtran}
\IEEEoverridecommandlockouts
\usepackage{cite}
\usepackage{amsmath,amssymb,amsfonts}
\usepackage{algorithmic}
\usepackage{graphicx}
\usepackage{textcomp}
\usepackage{xcolor}
\usepackage{ucs}
\usepackage[utf8x]{inputenc}
\usepackage{cite, amsfonts, amssymb, amsthm, bm, bbm, graphicx, relsize, multirow, booktabs, tikz,subfigure,soul}
\usepackage{stackrel}
\usepackage[american]{babel}
\usepackage[T1]{fontenc}
\usepackage{algorithmic, algorithm}
\usepackage[multiple]{footmisc}

\newcommand{\ds}{\displaystyle}
\newcommand{\norm}[1]{\left\lVert#1\right\rVert}

\def\BibTeX{{\rm B\kern-.05em{\sc i\kern-.025em b}\kern-.08em
    T\kern-.1667em\lower.7ex\hbox{E}\kern-.125emX}}
\begin{document}
\bstctlcite{IEEEexample:BSTcontrol}
\title{Using Massive MIMO Arrays for Joint Communication and Sensing}

\author{\IEEEauthorblockN{Stefano Buzzi\IEEEauthorrefmark{1}, Carmen D'Andrea\IEEEauthorrefmark{1} and Marco Lops\IEEEauthorrefmark{2}}
\IEEEauthorblockA{\IEEEauthorrefmark{1}\textit{DIEI - University of Cassino and Southern Latium},
Cassino, Italy. \\ Emails: \{buzzi,carmen.dandrea\}@unicas.it}
\IEEEauthorblockA{\IEEEauthorrefmark{2} \textit{DIETI - University "Federico II" of Naples}, 
Naples, Italy. \\ Email:  lops@unina.it}}

\maketitle

\begin{abstract}
One of the trends that is gaining more and more importance in the field of beyond-5G and 6G wireless communication systems is the investigation on systems that jointly perform communication and sensing of the environment. This paper proposes to use a base station (BS), that we call \textit{radar-BS}, equipped with a large-scale antenna array to execute, using the same frequency range, communication with mobile users and sensing/surveillance of the surrounding environment through radar scanning. The massive antenna array can indeed both operate as a MIMO radar with co-located antennas -- transmitting radar signals pointing at positive elevation angles -- and perform signal-space beamforming to communicate with users mainly based on the ground. Our results show that using a massive MIMO radar-BS the communication and the radar system can coexist with little mutual interference. 
\end{abstract}


\section{Introduction}
Recently, the coexistence between radar and communication systems and the consideration of systems that jointly perform communication and sensing of the environment are topics that have been attracting increasing interest in the field of beyond-5G wireless communication systems \cite{ZhengSPM,Aubry2015Optimizing,JSTSP,DAndrea_TWC2019}.  Indeed, since frequencies used by wireless communication systems are scaling up and trespassing frequency ranges used by radar systems, researchers have started to investigate methods to enable co-existence and spectrum sharing  between radar systems and communications systems \cite{ZhengSPM}. 
Within this research area, relevant efforts have been focused on the case in which the communication systems and the radar system share the same frequency range but are not co-located; in particular, the case in which the two systems cooperate to achieve a jointly optimal operating point and the case in which selfish design rules are employed have been both analyzed. Recently, the paper \cite{DAndrea_TWC2019} has analyzed the effect that radar interference has on the uplink of a wireless network wherein the base station (BS) is equipped with a large scale antenna array, showing that in the limit of large number of antennas at the BS, under some conditions, the radar interference to the communication system can be strongly reduced.

This paper considers instead the case in which the radar and communication function are co-located in a radar-BS and use the same massive antenna array\footnote{A similar scenario has been considered in \cite{Caire_OTFS_ICC2019} with reference to a vehicular radar and communication system.}. 
The motivation for considering such an architecture is at least twofold. On one hand, the use of large-scale antenna arrays and of digital beamforming makes it possible to simultaneously steer narrow multiple beams towards different positions. It thus follows that the radar-BS antenna array can communicate with the users, generally located  on the ground or at moderate heights, using signal-space beamforming, and, simultaneously, accomplish surveillance tasks of the surrounding area transmitting beams with positive elevation angles. On the other hand, the technological progress we are witnessing is such that in the near future several unmanned flying objects will populate the sky above our heads, and it thus becomes critical to be able to safely control and track them. The large arrays BSs are equipped with represent a precious resource that can be readily used to build with little efforts a network of radars aimed at short-to-medium range sky surveillance in urban areas.

This paper is organized as follows. Next section contains the description of the considered scenario and of the signal model. Section III is devoted to the discussion of the transceiver algorithms used at the BS for both communications and surveillance tasks. Performance analysis is carried out in Section IV, along with the discussion of the numerical results, while, finally, concluding remarks are given in Section V.

\section{System model}
\begin{figure}
\begin{center}
\includegraphics[scale=1.3]{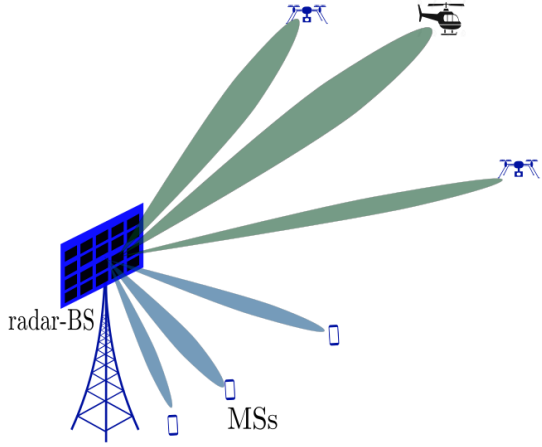}
\end{center}
\caption{Representation of the considered scenario.}
\label{Fig:scenario}
\end{figure}

We consider the scenario depicted in Fig. \ref{Fig:scenario}. A radar-BS equipped with a large-scale planar antenna array with $N_A=N_{A,y}N_{A,z}$ elements  ($N_{A,y}$ on the horizontal axis and $N_{A,z}$ on the vertical axis), jointly serves $K$ single-antenna mobile stations and performs surveillance tasks of the surrounding space -- through electronically steered phased-array beams pointed at  positive elevation angles -- using the same frequency range. The time-division-duplex (TDD) protocol is used for data communication with the mobile stations, so as to exploit the uplink/downlink channel reciprocity.
We denote by $B$ the total bandwidth and by $f_c$ the carrier frequency. Orthogonal 
frequency division multiplexing (OFDM) modulation is used for both communication and surveillance tasks; the total bandwidth is thus divided into $M$ subcarriers, i.e. $B=M \Delta f$, where $\Delta f$ denotes the subcarrier bandwidth.

\subsection{Channel model} \label{Channel_model_section}
We now provide the model for the channel between the BS and the potential target. 
Assume that a target with radial speed  $v$ [m/sec] with respect to the radar-BS is present in the surveilled area. 
The channel from the BS to the target and then, upon reflection, again to the BS is modeled as a random linear time variant system with matrix-valued channel impulse response expressed as 
\begin{equation}
\widetilde{\mathbf{H}}_T(t,\tau)=\mathbf{H}_T \delta(t-\tau)e^{j2\pi\nu t} \; .
\label{radar_channel}
\end{equation}
In \eqref{radar_channel}, 
$\tau$ and $\nu$ denote the round-trip delay and the Doppler shift induced by the target speed; moreover, letting the pair $(\phi, \theta)$ denote the azimuth and elevation angles of the target with respect to the BS antenna, we have 
$
\mathbf{H}_T= \alpha_T \mathbf{a}\left(\phi,\theta\right) \mathbf{a}\left(\phi,\theta\right)^H,
$
with $\alpha_T$ a complex coefficient taking into account the  target reflection coefficient and the path-loss. The vector  $\mathbf{a}\left(\phi,\theta\right)$ represents the BS antenna array response vector associated with the angles $(\phi, \theta)$, i.e.,
\begin{equation}
\begin{array}{lllll}
\mathbf{a}\left(\phi,\theta\right)&=\left[1, \ldots, e^{-j\tilde{k}d\left(a_y \sin(\phi)\sin(\theta) + a_z \cos(\theta)\right)}, \right. \\ & \left. \ldots, e^{-j\tilde{k}d\left((N_{A,y}-1) \sin(\phi)\sin(\theta) + (N_{A,z}-1) \cos(\theta) \right)} \right] 
\end{array}
\label{array_response}
\end{equation}
with $\tilde{k}=2\pi/\lambda$  the wavenumber, $\lambda$ the wavelength and $d$ the inter-element spacing.

With regard to the channel between the radar-BS and the generic $k$-th user,
$\mathbf{h}_k$ say, three different scenarios will be considered: Rayleigh-distributed channel, pure line-of-sight (LoS) channel with uniform phase, and Rice-distributed channels. For the Rayleigh case, we have
\begin{equation}
\mathbf{h}_k= \sqrt{\beta_k} \mathbf{g}_k \, ,
\label{Rayleigh_channel}
\end{equation}
where $\beta_k$ subsumes the path-loss and the shadow fading coefficient, and $\mathbf{g}_k \sim \mathcal{CN}\left(\mathbf{0}, \mathbf{I}_{N_A}\right)$. 
If the LoS channel model is in force, we have 
\begin{equation}
\mathbf{h}_k= \sqrt{\beta_k} e^{j \psi_k} \mathbf{a}\left(\varphi_k,\vartheta_k\right) \, ,
\label{LOS_channel}
\end{equation}
with $\beta_k$ representing the path-loss, $\psi_k$ is  the random phase uniformly distributed in $[0, 2\pi]$ and $\mathbf{a}\left(\varphi_k,\vartheta_k\right)$ is the BS antenna array response  evaluated at the azimuth and elevation angles, $\left(\varphi_k,\vartheta_k\right)$ say, of the $k$-th user.
Finally, for the  Rice-distributed  channel we have
\begin{equation}
\mathbf{h}_k= \sqrt{\frac{\beta_k}{K_{k}+1}} \left[ \sqrt{K_k} e^{j \psi_k} \mathbf{a}\left(\varphi_k,\vartheta_k\right) +\mathbf{g}_k \right]\, ,
\label{Rice_channel}
\end{equation}
where $K_k=\left[p_{\rm LoS}(d_{k, {\rm 2D}})\right]/ \left[1-p_{\rm LoS}(d_{k, {\rm 2D}})\right]
$ is the Ricean $K$-factor,
$d_{k, {\rm 2D}}$ is the 2D distance between the BS and the  $k$-th user, and $p_{\rm LoS}(d_{k, {\rm 2D}})$ is the LoS probability.

\subsection{Signal model: uplink channel estimation}
We now provide the signal model for the uplink channel estimation (CE) phase. Since the BS does not transmit during this phase, the received signal will not contain any possible target echo.
Let us denote by $\tau_c$ the dimension in time/frequency samples of the channel coherence length, and by $\tau_p < \tau_c$ the dimension of the uplink training phase. 
We also denote by $\boldsymbol{\phi}_k \in \mathbb{C}^{\tau_p}$ the pilot sequence transmitted by the $k$-th user, with $\|\boldsymbol{\phi}_k\|^2=1\, , \forall k$. 
Based on the above assumptions, the signal received at the radar-BS during the training phase 
 can be therefore expressed as the following $({N_{A}  \times \tau_p})$-dimensional matrix:
\begin{equation}
\mathbf{Y}_{\rm p} = \ds \sum_{k=1}^K \ds \sqrt{\eta_{{\rm p},k}} \mathbf{h}_{k}\boldsymbol{\phi}_k^H + \mathbf{W}_{\rm p} \; ,
\label{eq:y_pilot}
\end{equation}
with $\eta_{{\rm p},k}$ denoting the $k$-th user transmitted power, and $\mathbf{W}_{\rm p} \in \mathbb{C}^{N_A  \times \tau_p}$ represents the thermal noise contribution and out-of-cell interference at the radar-BS. The entries of  $\mathbf{W}_{\rm p}$  are modeles as i.i.d.  ${\cal CN}(0, \sigma^2_w)$ RVs.

\subsection{Signal model: downlink communication and radar function}

Consider now the downlink transmission phase. Following \cite{Caire_OTFS_ICC2019}, we assume that a standard cyclic prefix (CP) OFDM modulation is used for both the communication and radar surveillance tasks, with $\Delta f$ the subcarrier spacing. 
Let $T_0=T_{ \rm CP} +T_{\rm s}$ be the OFDM symbol duration, with $T_{ \rm CP}$ and $T_{\rm s}=1/\Delta f$ denoting the CP and the symbol duration, respectively. The OFDM frame duration is $T_{\rm OFDM}=N T_0$. The unit-power data symbols intended for the $k$-th user are denoted by $x_k(n,m)$ for $n=0,\ldots, N-1$,  $m=0,\ldots, M-1$, and  are arranged in a $N \times M$ grid. 
Similarly, the fictitious unit-power symbols used for radar detection are denoted by  $x_R(n,m)$ and arranged in a $N \times M$ grid. The continuous-time OFDM signals with CP intended to the $k$-th user  and intended for radar surveillance can be thus written as
\begin{equation}
s_k(t)=\ds \sum_{n=0}^{N-1} \sum_{m=0}^{M-1} x_k(n,m) \text{rect} (t-n T_0) e^{j2\pi m \Delta f (t- T_{\rm CP}-nT_0)},
\label{signal_k}
\end{equation}
and
\begin{equation}
s_R(t)=\ds \sum_{n=0}^{N-1} \sum_{m=0}^{M-1} x_R(n,m) \text{rect} (t-n T_0) e^{j2\pi m \Delta f (t- T_{\rm CP}-nT_0)},
\label{signal_R}
\end{equation}
respectively, with  $\text{rect}(t)$ a rectangular pulse supported on $[0, T_0]$. 
Accordingly,
denoting by $\eta_k$ the power used by the radar-BS to transmit to the $k$-th 
user\footnote{Uniform power allocation across users is used in this paper, i.e.
$\eta_k=P_{\rm DL}/(K M N)$, with $P_{\rm DL}$ the radar-BS power budget used for communication tasks.}, 
and letting
$\eta_R=P_R/(M N)$, with $P_R$ the power used for surveillance purposes, 
the $N_A$-dimensional signal transmitted by the radar-BS can be written as
\begin{equation}
\mathbf{s}(t)=\ds \sum_{k=1}^K { \sqrt{\eta_k} s_k(t) \mathbf{w}_k } + \sqrt{\eta_R} s_R(t) \mathbf{w}_R\left(\phi,\theta\right) \, .
\label{signal_DL_st}
\end{equation}
In \eqref{signal_DL_st},  $\mathbf{w}_k$ is the beamforming vector used to transmit to the $k$-user, while $\mathbf{w}_R\left(\phi,\theta\right)$ is beamforming vector for surveillance tasks in the  direction corresponding to the azimuth and elevation angles. 
Two possible choices are considered for $\mathbf{w}_R\left(\phi,\theta\right)$. 
The former is to use the radar-BS antenna as a phased array producing a phased beam towards the direction $(\phi,\theta)$, i.e.:
\begin{equation}
\mathbf{w}_R\left(\phi,\theta\right)= \frac{1}{\sqrt{N_A}}\mathbf{a}\left(\phi,\theta\right)\, .
\label{pure_darad_direction}
\end{equation}
The above choice would however provide some disturbance to ground users; an alternative is thus to modify the beamformer in \eqref{pure_darad_direction} in order to force to zero the interference produced by the radar signal to the mobile users. 
Letting $\widetilde{\mathbf{U}}$ denote a matrix whose columns form a basis for the subspace spanned by the estimated channel vectors $\left[  \widehat{\mathbf{h}}_1, \ldots,  \widehat{\mathbf{h}}_K \right]$, we have thus the zero-forcing radar (ZFR) beamformer:
\begin{equation}
\mathbf{w}_R\left(\phi,\theta\right)= \frac{\left( \mathbf{I}_{N_A}- \widetilde{\mathbf{U}} \widetilde{\mathbf{U}}^H\right) \mathbf{a}\left(\phi,\theta\right)}{\norm{\left( \mathbf{I}_{N_A}- \widetilde{\mathbf{U}} \widetilde{\mathbf{U}}^H \right) \mathbf{a}\left(\phi,\theta\right)}}\, .
\label{ZF_radar}
\end{equation}

\section{Transceiver algorithms}

\subsection{Uplink channel estimation}
Given the 
observable $\mathbf{Y}_{\rm p}$ reported in \eqref{eq:y_pilot}, the radar-BS forms the statistics 
$\mathbf{y}_{{\rm p},k}=\mathbf{Y}_{\rm p} \boldsymbol{\phi}_k$, $\forall \; k=1,\ldots, K$.
In order to estimate the channel vectors 
$\mathbf{h}_{k}, \forall \; k=1,\ldots, K$, two possible CE techniques will be considered:
 pilot matched CE (PM-CE) and  linear minimum-mean-square-error CE (LMMSE-CE). 
 
 For the case of PM-CE, the channel estimate of $\mathbf{h}_k$ is written as
$
 \widehat{\mathbf{h}}_k=\frac{1}{\sqrt{\eta_{{\rm p},k}}}\mathbf{y}_{{\rm p},k} \, .
$
For LMMSE-CE, instead, the channel estimate can be shown to be written as \cite{kay1993fundamentals} 
$
 \widehat{\mathbf{h}}_k=\mathbf{E}_k^H\mathbf{y}_{{\rm p},k} \, ,
$
where
$
\mathbf{E}_k= \sqrt{\eta_{{\rm p},k}} \mathbf{R}_{y,k}^{-1} \overline{\mathbf{H}}_k \, ,
$
$
\mathbf{R}_{y,k}= \sum_{i=1}^K \ds \eta_{{\rm p},i} \overline{\mathbf{H}}_i\left|\boldsymbol{\phi}_i^H \boldsymbol{\phi}_k \right|^2+ \sigma^2_w \mathbf{I}_{N_A} \, ,
$
and $\overline{\mathbf{H}}_k$ is an $(N_A \times N_A)$-dimensional matrix depending on the adopted channel model. For the case of Rayleigh-distributed channel, Eq. \eqref{Rayleigh_channel}, we have
$
\overline{\mathbf{H}}_k= \beta_k \mathbf{I}_{N_A} \, ;
$
for the case of LoS channel, Eq. \eqref{LOS_channel}, we have
$
\overline{\mathbf{H}}_k= \beta_k \mathbf{a}\left(\varphi_k,\vartheta_k\right) \mathbf{a}^H\left(\varphi_k,\vartheta_k\right) \, ,
$
while finally, for the case of Rice-distributed channel, Eq. \eqref{Rice_channel}, we have
\begin{equation}
\overline{\mathbf{H}}_k= \frac{\beta_k}{K_{k}+1}\left[ K_{k} \mathbf{a}\left(\varphi_k,\vartheta_k\right) \mathbf{a}^H\left(\varphi_k,\vartheta_k\right) + \mathbf{I}_{N_A} \right]\, .
\label{H_bar_matrix_Rice}
\end{equation}

\subsection{Radar processing}

Let us now focus on the derivation of the signal processing tasks for the radar, whose aim  is to discriminate between the hypothesis $H_0$, no target, and the hypothesis $H_1$, a target is present, for any scanned range-cell. 
In order to perform joint radar detection in the direction defined by the angles $(\phi, \theta)$ and communication with the users, the following discrete-time signal is transmitted  
\begin{equation}
\begin{array}{llll}
\mathbf{s}(n,m)=\left[ \!\ds  \sum_{k=1}^K { \!\!\!\sqrt{\eta_k} x_k(n,m) \mathbf{w}_k } \!\!   + \!\!\sqrt{\eta_R} x_R(n,m) \mathbf{w}_R\left(\phi,\theta\right)\!\!\right]\, .
\end{array}
\label{DL_signal}
\end{equation}
Assuming that a target is present , the following signal is received at the radar-BS:
\begin{equation}
\widetilde{\mathbf{y}}(t)=\mathbf{H}_T \mathbf{s}(t-\tau)e^{j2\pi\nu t} + \widetilde{\mathbf{z}}(t),
\label{received_signal}
\end{equation}
with $\tau$ the delay induced by the target distance and $\nu$ the doppler frequency offset induced by the target radial speed. 
Sampling the received waveform  every $T_{\rm s}/M$ and removing  the CP removed in each OFDM symbol we obtain \cite{Caire_OTFS_ICC2019}:
\begin{equation}
\begin{array}{llll}
\widetilde{\mathbf{y}}(n,m)=&\!\!\!\!\mathbf{H}_T e^{j2\pi\nu n T_0}\ds \!\!\!\!\sum_{\ell=0}^{M-1} \mathbf{s}(n,\ell) \ds e^{  j2\pi\frac{m}{M} \left( \frac{\nu}{\Delta f} + \ell \right)} e^{ -j2\pi\ell \Delta f \tau}  \\ &+\widetilde{\mathbf{z}}(n,m).
\end{array}
\end{equation}
Applying the discrete Fourier transform (DFT) and exploiting the orthogonal property, the received signal is transformed as
\begin{equation}
\begin{array}{llll}
&\mathbf{y}(n,m)= \!\!\ds \frac{1}{M} \sum_{q=0}^{M-1} \!\!{\widetilde{\mathbf{y}}(n,q)} e^{-j2 \pi \frac{mq}{M}}   \\ & \approx \mathbf{H}_T e^{j2\pi\nu n T_0} e^{-j2\pi m \Delta f \tau}  \ds \mathbf{s}(n,m) + \mathbf{z}(n,m),
\end{array}
\label{output_H1}
\end{equation}
where $\mathbf{z}(n,m)$ is the DFT of the noise contribution and the approximation follows by letting $\nu_{\rm max} \ll \Delta f$.

If, instead, the target is absent, the output is simply written as $\mathbf{y}(n,m)=\mathbf{z}(n,m)$.
We thus formulate the detection problem as the following binary hypothesis test
\begin{equation}
\left\{\!\begin{array}{llll}
H_1: & \!\! \mathbf{y}(n,m)=\alpha_T\mathbf{u}(n,m) e^{j2\pi\nu n T_0} e^{-j2\pi m \Delta f \tau} \!+ \mathbf{z}(n,m) \\
H_0: & \!\! \mathbf{y}(n,m)=\mathbf{z}(n,m)  \, ,
\end{array} \right.
\label{HT}
\end{equation}
with 
$
\mathbf{u}(n,m) = \mathbf{a}\left(\phi,\theta\right) \mathbf{a}\left(\phi,\theta\right)^H \mathbf{s}(n,m)\, .
$
Given the total ignorance on the parameters $\alpha_T, \nu, \tau$, 
upon defining the uniformly-spaced grid in the delay and Doppler domain $\mathcal{G}$, the Generalized Likelihood Ratio Test (GLRT) can be implemented as follows 
\begin{equation}
\max_{\tau, \nu \in \cal{G}} \left|  \ds \sum_{n=0}^{N-1} \sum_{m=0}^{M-1} e^{-j2\pi\nu n T_0} e^{j2\pi m \Delta f \tau} \mathbf{u}(n,m)^H \mathbf{y}(n,m) \right|^2\stackrel[H_0]{H_1}{\gtrless} \gamma
\label{GLRT}
\end{equation}
\subsection{Downlink processing}
On the downlink, the signal received by the $k$-th user  is expressed in discrete-time as follows:
\begin{equation}
\begin{array}{llll}
y_k(n,m)=&\sqrt{\eta_k} \mathbf{h}_k^H \mathbf{w}_k x_k(n,m) + \ds \sum_{\substack{j=1 \\ j \neq k}}^K {\sqrt{\eta_j} \mathbf{h}_k^H \mathbf{w}_j x_j(n,m)}  \\& + \sqrt{\eta_R} \mathbf{h}_k^H \mathbf{w}_R\left(\phi,\theta\right) x_R(n,m) + z_k(n,m) \, ,
\end{array}
\label{DL_signal_user}
\end{equation}
where $z_k(n,m)\sim {\cal CN}(0, \sigma^2_z)$ is the AWGN contribution. The quantity $y_k(n,m)$ thus represents the soft estimate of the information symbol $x_k(n,m)$ and can be further processed for data detection. 
\begin{figure*}
\begin{equation}
\begin{array}{ll}
\mathcal{R}_k^{\rm PM}= \ds \frac{\tau_d}{\tau_c}\log_2 \left( 1+
\ds \frac{\eta_k \gamma_k}
{\ds \sum_{j=1}^K {\frac{\eta_j}{\eta_{{\rm p},j}} \left( \frac{\text{tr}\left( \mathbf{R}_{y,j} \overline{\mathbf{H}}_k \right) }{\gamma_j} + \eta_{{\rm p},k} \frac{\delta_k}{\gamma_j} \left|\boldsymbol{\phi}_k^H \boldsymbol{\phi}_j \right|^2\right) } - \eta_k \gamma_k +\eta_R \mathbf{w}_R^H\left(\phi,\theta\right) \overline{\mathbf{H}}_k \mathbf{w}_R\left(\phi,\theta\right) + \sigma^2_z }\right) 
\end{array}
\label{SE_DL_PM}
\end{equation}
\begin{equation}
\begin{array}{ll}
\mathcal{R}_k^{\rm LMMSE}= \ds \frac{\tau_d}{\tau_c}\log_2 \left( 1+
\ds \frac{\eta_k \widetilde{\gamma}_k}
{ \ds \sum_{j=1}^K {\eta_j \left(\sqrt{\eta_{{\rm p},j}} \frac{\text{tr}\left(  \overline{\mathbf{H}}_j\mathbf{E}_{j} \overline{\mathbf{H}}_k \right)}{\widetilde{\gamma}_j}  + \eta_{{\rm p},k} \frac{\widetilde{\delta}_j^{(k)}}{\widetilde{\gamma}_j} \left|\boldsymbol{\phi}_k^H \boldsymbol{\phi}_j \right|^2 \right) } - \eta_k \widetilde{\gamma}_k +\eta_R \mathbf{w}_R^H\left(\phi,\theta\right) \overline{\mathbf{H}}_k \mathbf{w}_R\left(\phi,\theta\right)+ \sigma^2_z }\right) 
\end{array}
\label{SE_DL_MMSE}
\end{equation}
\hrulefill
\end{figure*}

\section{Performance measures}
Regarding the radar, the used performance measures are the usual probability of detection and probability of false alarm. 

With regard, instead, to the downlink communication system, the downlink achievable rates are taken as performance measure.
In particular, starting from Eq. \eqref{DL_signal_user}, and exploiting the use-and-then-forget bounding technique \cite{marzetta2016fundamentals},  the closed form achievable rate formulas, reported 
in Eqs. \eqref{SE_DL_PM} and \eqref{SE_DL_MMSE} at the top of next page, can be derived 
for the PM-CE and for the LMMSE-CE, assuming channel matched beamforming, i.e., $\mathbf{w}_k=\widehat{\mathbf{h}}_k/\norm{\widehat{\mathbf{h}}_k}$, respectively. In these formulas,  $\tau_d=\tau_c-\tau_p$ is the dimension in time/frequency samples of the downlink data transmission phase, 
$
\gamma_k=\text{tr}\left(\overline{\mathbf{H}}_k\right)$, and $\widetilde{\gamma}_k=\sqrt{\eta_{{\rm p},k}}\text{tr}\left(\overline{\mathbf{H}}_k\mathbf{E}_k\right).
$ Moreover, for the case of Rayleigh channel, we have
$
\delta_k= \beta_k^2 N_A^2 \; \; \text{and} \; \; \widetilde{\delta}_j^{(k)}=\beta_k^2\text{tr}\left(\mathbf{E}_j^H\right) 
$;  for the case of LoS channel, we have
$
\delta_k= 0 \; \; \text{and} \; \; \widetilde{\delta}_j^{(k)}=0
\label{delta_LOS}
$;
and, finally, for the case of Rice channel, we have
$
\delta_k= \left( \frac{\beta_k}{K_{k}+1} \right)^2 N_A \left( N_A+ 2 K_{k}\right)
\label{delta_k_Rice}
$, and 
$\widetilde{\delta}_j^{(k)}=  \left(  \frac{\beta_k}{K_{k}+1} \right)^2 \left[\text{tr}\left( \mathbf{E}_j^H \right)  \right. $ $ \left.+ \ds  2 K_{k} \mathbb{R}\left\lbrace  \text{tr}\left( \mathbf{a}^H\left(\varphi_k,\vartheta_k\right)\mathbf{E}_j^H \mathbf{a}\left(\varphi_k,\vartheta_k\right)  \mathbf{E}_j\right)\right\rbrace  \right].
$

\begin{table}[]
\centering
\caption{Simulation Parameters}
\label{Sim_Par}
\begin{tabular}{|p{1.5cm}|p{1.5cm}|p{4cm}|}
\hline
\textbf{Name}                   & \textbf{Value} & \textbf{Description}                                                                                                                \\ \hline
$f_c$                           & 3 GHz          & carrier frequency                                                                                                                   \\ \hline
$M$                             & 512    & number of subcarriers
 \\ \hline
$N$                             & 14    & number of OFDM symbols
 \\ \hline
$\Delta_f$                            & 30 kHz    & subcarrier spacing
 \\ \hline
 $B=\Delta_f M$                            & 15.36 MHz    & system bandwidth
 \\ \hline
  $T_0$                            & 0.357 $\mu$s    & OFDM symbol duration
 \\ \hline
$K$                             & 10          & number of users in the cellular system \\ \hline
$P_{\rm DL}$                           &  2 W         & Power used for downlink communication 
\\ \hline
$F$                             & 9 dB           & noise figure at the receiver                                                                                                        \\ \hline
$\mathcal{N}_0$                 & -174 dBm/Hz    & power spectral density of the noise                                                                                                 \\ \hline
\end{tabular}
\end{table}

\begin{figure*}[!t]
\centering
\includegraphics[scale=0.4]{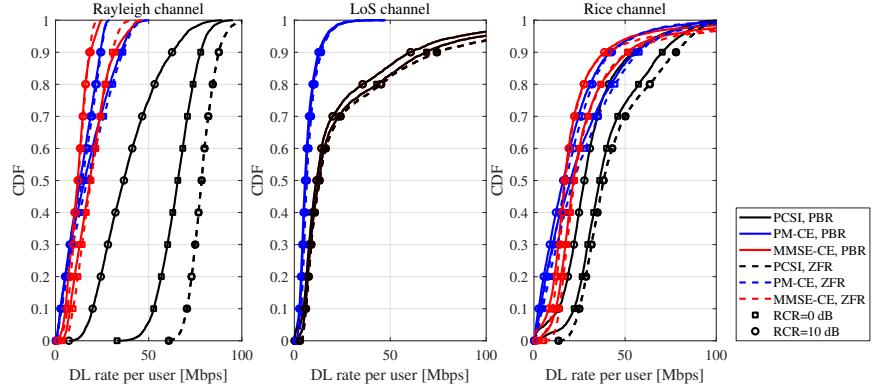}
\caption{CDFs of DL rate per user using the PBR and ZFR approaches. Rayleigh channel, LoS channel and Rice channel, $N_{A,y}\times N_{A,z}=20 \times 20$, two values of RCR.}
\label{Fig:Rate_channels_PBR_ZFR}
\end{figure*}

\section{Numerical results}

The parameters for the simulation setup are reported in Table \ref{Sim_Par}. We assume that the users of the communication system are randomly located on the $(x,y)$ plane with $x$ in $[10,100]$ m and $y$ in $[-50,-10] \cup [10, 50]$, with heighs 1.65 m. The height of the radar-BS is 15 m. 
For the Rayleigh channel model in Eq. \eqref{Rayleigh_channel}, we follow the three slope path loss model in \cite{buzzi_CFUC2017} and we assume uncorrelated shadow fading. For the LoS channel in Eq. \eqref{LOS_channel}, the path-loss follows the model in \cite[Table B.1.2]{3GPP_36814_GUE_model}, while  for the Rice channel in Eq. \eqref{Rice_channel} we use again the model in \cite{buzzi_CFUC2017} and  the LoS probability is evaluated following \cite{3GPP_36873}. The quantity $\alpha_T$ in Eq. \eqref{radar_channel} containing the target reflection coefficient and the path-loss is modeled as
$\alpha_T=G\sqrt{\frac{\zeta}{L_{\tau}}}$, 
where $G=10 \log_{10}(N_A)$ dB is the radar-BS antenna gain, $\zeta=0.1253 \text{m}^2$ is the target radar cross section (RCS)\footnote{The RCS of a common unmanned aherial vehicle (UAV) \cite{Chenchen_2016_Drones} has been chosen.} and 
$L_{\tau}=\frac{(4 \pi)^3}{\lambda^2}\left( \frac{c \tau}{2}\right)^4$.
We define the Radar-Communication-Ratio (RCR) as  
$
\text{RCR}=P_R/P_{\rm DL}.
$
The scanning area of the radar system  extends for $[-60, 60]^o$ in azimuth and for $[10, 80]^o$ in elevation.

\begin{figure}[!t]
\centering
\includegraphics[scale=0.3]{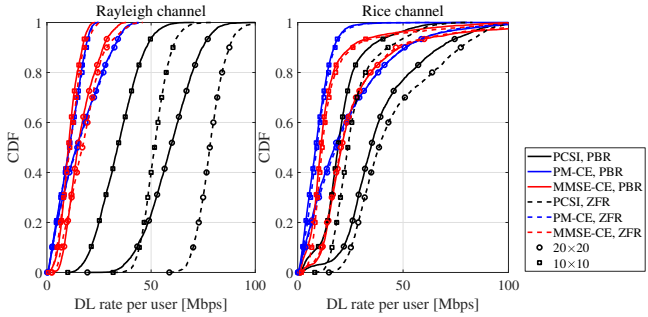}
\caption{CDF of DL rate per user using the PBR and ZFR approaches. Rayleigh and Rice channels, RCR $=3$ dB, two values of $N_{A,y}\times N_{A,z}$.}
\label{Fig:Rate_AntennaConf_PBR_ZFR}
\end{figure}

\begin{figure}[!t]
\centering
\includegraphics[scale=0.26]{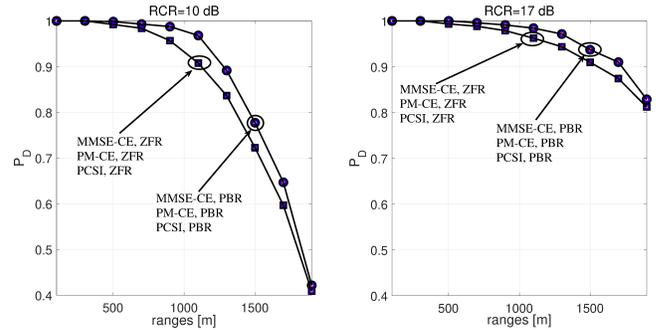}
\caption{Probability of detection versus range using the PBR and ZFR approaches. Rayleigh channel for the users, two values of RCR, $N_{A,y}\times N_{A,z}=10 \times 10$.  }
\label{Fig:Detection_probability}
\end{figure}

Fig. \ref{Fig:Rate_channels_PBR_ZFR} reports the cumulative distribution functions (CDFs) of the downlink (DL) rate per user for the three channel models discussed in Section \ref{Channel_model_section} and for two values of RCR. 
Results show that the presence of the radar system has some effect on the users achievable rates, even though the use of ZFR beamforming helps t restore the system performance. The figure also permits assessing the impact of the CE techniques on the system performance. 
Fig. \ref{Fig:Rate_AntennaConf_PBR_ZFR} reports the DL rate per user in the cases of Rayleigh and Rice channels, for fixed RCR,  and for two values of antenna configurations at the radar-BS. The figure permits assessing the beneficial impact of the increase of the antenna array size. 
In Fig. \ref{Fig:Detection_probability}, we report the probability of detection $P_D$ versus the target distance, using Rayleigh channel for the users and two values of RCR. The threshold $\gamma$ in Eq. \eqref{GLRT} has been numerically evaluated assuming a false alarm probability of $10^{-2}$. It can be seen  that the performance in the case of CE and PCSI is the same, due to the fact that the knowledge of the channels between radar-BS and users does not play any role in the detection capability of the system. Additionally, as expected, it is shown  that the detection performance in the case of ZFR is worse than that achieved with  PBR: indeed,  nulling the interference between the radar signal and the users has a negative impact on the shape of the beam used for target detection.

\section{Conclusions}
The paper has analyzed the case in which a radar-BS equipped with massive MIMO arrays is used for joint communications and sensing tasks. A system model for the proposed architecture  and the related signal processing algorithms have been derived.  Promising numerical results have been shown, which are a ground for further investigations on this subject. 

\ifCLASSOPTIONcaptionsoff
  \newpage
\fi
\bibliographystyle{IEEEtran}

\bibliography{MyReference}

\end{document}